\documentclass[sigconf]{acmart}

\usepackage[utf8]{inputenc}
\usepackage{booktabs}
\usepackage{adjustbox}
\usepackage{graphicx}
\usepackage{subcaption}
\usepackage{caption}
\usepackage{multirow}
\usepackage{enumerate}
\usepackage{enumitem}
\usepackage{amsmath}
\usepackage{graphicx}
\usepackage{amsfonts}
\usepackage{amssymb}
\usepackage{url}
\usepackage{acronym}
\usepackage{color}
\usepackage{booktabs}
\usepackage{diagbox}

\usepackage[T1]{fontenc}
\usepackage[font=small,labelfont=bf,tableposition=top]{caption}

\DeclareCaptionLabelFormat{andtable}{#1~#2  \&  \tablename~\thetable}

\usepackage{array}
\newcolumntype{L}[1]{>{\raggedright\let\newline\\\arraybackslash\hspace{0pt}}m{#1}}
\newcolumntype{C}[1]{>{\centering\let\newline\\\arraybackslash\hspace{0pt}}m{#1}}
\newcolumntype{R}[1]{>{\raggedleft\let\newline\\\arraybackslash\hspace{0pt}}m{#1}}

\copyrightyear{2018} 
\acmYear{2018} 
\setcopyright{acmlicensed}
\acmConference[CIKM '18]{The 27th ACM International Conference on Information and Knowledge Management}{October 22--26, 2018}{Torino, Italy}
\acmBooktitle{The 27th ACM International Conference on Information and Knowledge Management (CIKM '18), October 22--26, 2018, Torino, Italy}
\acmPrice{15.00}
\acmDOI{10.1145/3269206.3269301}
\acmISBN{978-1-4503-6014-2/18/10}

\begin{document}

\title{Improved and Robust Controversy Detection in General Web Pages Using Semantic Approaches under Large Scale Conditions}
\author{Jasper Linmans}
\affiliation{%
  \institution{University of Amsterdam}
}
\email{jasperlinmans@gmail.com}
\author{Bob van de Velde}
\affiliation{%
  \institution{University of Amsterdam}
}
\email{rnvandevelde@gmail.com}
\author{Evangelos Kanoulas}
\affiliation{%
  \institution{University of Amsterdam}
}
\email{ekanoulas@gmail.com}

\renewcommand{\shortauthors}{J. Linmans et al.}
\renewcommand{\shorttitle}{}

\begin{abstract}
Detecting controversy in general web pages is a daunting task, but increasingly essential to efficiently moderate discussions and effectively filter problematic content. Unfortunately, controversies occur across many topics and domains, with great changes over time. This paper investigates neural classifiers as a more robust methodology for controversy detection in general web pages. Current models have often cast controversy detection on general web pages as Wikipedia linking, or exact lexical matching tasks. The diverse and changing nature of controversies suggest that semantic approaches are better able to detect controversy. We train neural networks that can capture semantic information from texts using weak signal data. By leveraging the semantic properties of word embeddings we robustly improve on existing controversy detection methods. To evaluate model stability over time and to unseen topics, we asses model performance under varying training conditions to test cross-temporal, cross-topic, cross-domain performance and annotator congruence. In doing so, we demonstrate that weak-signal based neural approaches are closer to human estimates of controversy and are more robust to the inherent variability of controversies. 
\end{abstract}

%
%



\maketitle

\section{Introduction \& Prior work}

Controversy detection is an increasingly important task. Controversial content can signal the need for moderation on social platforms, either to prevent conflict  between users or limit the spread of misinformation. More generally, controversies provide insight into societies \cite{dori15}. Often, the controversial content is outside the direct control of a platform on which it is shared, mentioned or discussed. This raises the requirement of generally applicable methods to gauge controversial content on the web for moderation purposes. Unfortunately, what is controversial changes, and may lie more in the \textit{way} topics are discussed rather than \textit{what} is discussed, making it difficult to detect controversies in a robust fashion. We take the task of controversy detection and evaluate robustness of different methodologies with respect to the varying nature of controversies.

Prior work on detecting controversies has taken three kinds of approaches: 1) \textbf{lexical approaches}, which seek to detect controversies through \textit{signal terms}, either through bag-of-word classifiers, lexicons, or lexicon based language models \cite{jang16}. 2) \textbf{explicit modeling} of controversy through platform-specific features, often in Wikipedia or social-media settings. Features such as mutual reverts \cite{yasseri12}, user-provided flags \cite{das15}, interaction networks \cite{popescu10} or stance-distributions \cite{jang2017modeling} have been used as platform-specific indicators of controversies. The downside of these approaches is the lack of generalizability due to their platform-specific nature. 3) \textbf{matching models} that combine lexical and explicit modelling approaches by looking at lexical similarities between a given text and a set of texts in a domain that provides explicit features \cite{ jang16, Dori13, jang16b}.  

Controversy detection is a difficult task because 1) controversies are \textit{latent}, like ideology, meaning they are often not directly mentioned as controversial in text. 2) Controversies occur across a vast range of topics with varying topic-specific vocabularies. 3) Controversies change over time, with some topics and actors becoming controversial whereas others stop to be so.  Previous approaches lack the power to deal with such changes. Matching and explicit approaches are problematic when the source corpus (e.g. Wikipedia) lags after real-world changes \cite{grauss}. Furthermore, lexical methods trained on common (e.g. fulltext) features are likely to memorize the controversial topics in the training set rather than the `language of controversy'. Alleviating dependence on platform specific features and reducing sensitivity to an exact lexical representation is paramount to robust controversy detection. To this end, we focus only on fulltext features and suggest to leverage the semantic representations of word embeddings to reduce the vocabulary-gap for unseen topics and exact lexical representations.

The majority of NLP-task related neural architectures rely on word embeddings, popularized by Mikolov et al \cite{Mikolov13} to represent texts. In essence these embeddings are latent-vector representations that aim to capture the underlying meaning of words. Distances between such latent-vectors are taken to express semantic relatedness, despite having different surface forms. By using embeddings, neural architectures are also able to leverage features learned on other texts (e.g. pretrained word embeddings) and create higher level representations of input (e.g. convolutional feature maps or hidden-states). These properties suggest that neural approaches are better able to generalize to unseen examples that poorly match the training set. We use two often applied network architectures adopting word embeddings, to classify controversy: Recurrent Neural Networks \cite{Yang16} and Convolutional Neural Networks \cite{Kim14} to answer the following research question. 
\textbf{RQ:} Can we increase robustness of controversy detection using neural methods?

Currently, there is no open large-size controversy detection dataset that lends itself to test cross-temporal and cross-topic stability. Thus we generate a Wikipedia crawl-based dataset that includes general web pages and is sufficiently large to train and test high capacity models such as neural networks. 




\section{Methods}
A proven approach in modelling text with neural networks is to use Recurrent Neural Networks (RNNs) which enjoy weight sharing capabilities to model words irrespective of their sequence location. A specific type, the Hierarchical Attention Network (HAN) proposed by \cite{Yang16} makes use of attention to build document representations in a hierarchical manner. It uses bi-directional Gated Recurrent Units (GRUs) \cite{Bahdanau14} to selectively update representations of both words and sentences. This allows the network to both capture the hierarchy from words to sentences to documents and to explicitly weigh all parts of the document relevant during inference. 

Recently, Convolutional Neural Networks (CNNs) have enjoyed increasing success in text classification. One such network introduced by \cite{Kim14} looks at patterns in words within a window, such as "Scientology [...] brainwashes people". The occurrences of these patterns are then summarized to their 'strongest' observation (max-pooling) and used for classification. Since pooling is applied after each convolution, the output size of each convolutional operation itself is irrelevant. Therefore, filters of different sizes can be used, each capturing patterns in different sized word windows. 

We explore the potential of RNNs and CNNs for controversy detection using both the HAN \cite{Yang16} and the CNN \cite{Kim14} model\footnote{Code available at https://github.com/JasperLinmans/ControversPy}. Similar to \cite{Yang16}, each bi-directional GRU cell is set to a dimension of 50, resulting in a word/sentence representation of size 100 after concatenation. The word/sentence attention vectors similarly contain 100 dimensions, all randomly initialized. The word windows defined in the CNN model are set to sizes: 2, 3 and 4 with 128 feature maps each. Each model is trained using mini batches of size 64 and uses both dropout (0.5) and $l_2$ regularization (1e-3) at the dense prediction layer. Both networks use pre-trained embeddings, trained on 100 billion words of a Google News corpus\footnote{Available at https://code.google.com/p/word2vec/}, which are further fine-tuned during training on the controversy dataset. The optimization algorithm used is Adam\cite{Kingma2014-ll} (learning rate: 1e-3).

\section{Experimental Setup}
\subsection{Datasets and evaluation}
We use the Clueweb09 derived dataset of \cite{dori15} for baseline comparison. For cross-temporal, cross-topic and cross-domain  training \& evaluation, we generate a new dataset based on Wikipedia crawl data\footnote{Script to generate dataset available at: https://github.com/JasperLinmans/ControversPy}. This dataset is gathered by using Wikipedia's   `List of Contoversial articles' overview page of 2018 (time of writing) and 2009 (for comparison with baselines) \footnote{Available at https://en.wikipedia.org/wiki/Wikipedia:List\_of\_controversial\_issues}. Using this as a `seed' set of controversial articles, we iteratively crawl the `See also', `References' and `External links' hyperlinks up to two hops from the seed list. The negative seed pages (i.e. non controversial) are gathered by using the random article endpoint\footnote{Available at https://en.wikipedia.org/wiki/Special:Random}. The snowball-sample approach includes general, non-Wikipedia, pages that are referred to from Wikipedia pages. The dataset thus extends beyond just the encyclopedia genre of texts. Labels are assumed to propagate: a page linked from a controversial issue is assumed to be controversial. The resulting dataset statistics are summarized in Table \ref{dataset}. 

\begin{table}[!htb]
\centering
\setlength{\tabcolsep}{6pt} 
\caption{Wikipedia derived dataset statistics. {\normalfont Including the percentages of controversial (i.e. positive labelled) and general (i.e. non-Wikipedia) web pages from the total amount of pages per dataset split.}}
\label{dataset}
\begin{tabular}{lcccc}
\toprule
 Set & Seeds & Total & Controversial & General Web\\
 \midrule
 Train & 5600 & 23.703 & 7.233 (31\%) & 15.449 (65\%)\\ 
 Validation & 200 & 988 & 651 (66\%) & 688 (70\%)\\ 
 Test & 200 & 1.024 & 654 (64\%) & 723 (71\%) \\
\bottomrule
\end{tabular}
\end{table}

To be useful as a flagging mechanism for moderation, a controversy detection algorithm should satisfy both Precision and Recall criteria. F1 scores will therefore be used to evaluate this balance. The AUC values are used to measure classification performance in the unbalanced controversy datasets. The test-train split depends on the task investigated and is listed in the results section for the respective task. To test for significant results, all models were evaluated using a bootstrap approach: by drawing 1000 samples with replacements $n$ documents from the test set equal to the test-set size. The resulting confidence intervals based on percentiles provide a measure of significance.




\subsection{Baseline models}
To compare the results of neural approaches to prior work we implemented the previous state-of-the-art controversy detection method: the language model from \cite{jang16b}. Together with an SVM baseline they act as controversy detection alternatives using only full text features, thus meeting the task-requirements of platform-independence. Note: the implementation of \cite{jang16b} additionally requires ranking methods to select a subset of the training data for each language model. A simplified version of this, excluding the ranking method but using the same dataset and lexicon to select documents as \cite{jang16b}, is implemented and included in the baselines comparison section (LM-DBPedia). We also included the same language model trained on the full text Wikipedia pages (LM-wiki). Similarly, for completeness sake, we also include both the state-of-the-art matching model, the TILE-Clique model from \cite{jang16} and the sentiment analysis baseline (using the state-of-the-art Polyglot library for python\footnote{https://github.com/aboSamoor/polyglot}) from \cite{Dori13} in the comparison with previous work.

\begin{figure*}
  \begin{minipage}{.70\textwidth}
\begingroup
\begin{table}[H]
\centering
\setlength{\tabcolsep}{1.6pt} 
\label{time-table}
\renewcommand{\arraystretch}{1.2} 
\vspace*{0.15cm}\hspace*{-0.75cm}\begin{tabular}{lcccccccccccc}
\toprule
Model                    &  \multicolumn{3}{c}{Precision}            & \multicolumn{3}{c}{Recall}            & \multicolumn{3}{c}{F1}                         & \multicolumn{3}{c}{AUC} \\ 
  $\ \ \ $\textit{Train/Test:} & \textit{'18/'18} & \textit{'09/'18} & $\Delta$ & \textit{'18/'18} & \textit{'09/'18} & $\Delta$& \textit{'18/'18} & \textit{'09/'18} & $\Delta$ & \textit{'18/'18} & \textit{'09/'18} & $\Delta$\\ 
\midrule
 TfIdf-SVM & 0.910  & \textbf{0.941} & $\blacktriangle 3\%$ & 0.689 & 0.191 & $\blacktriangledown 72\%$ & 0.784 & 0.317 & $\blacktriangledown 60\%$ & 0.785 & 0.585 & $\blacktriangledown 25\%$\\
 LM  &      0.651     &  0.609 & $\blacktriangledown 6\%$ &           0.811         & 0.550  & $\blacktriangledown 32\%$ & 0.723 & 0.578 & $\blacktriangledown 20\%$& 0.600 & 0.452 & $\blacktriangledown 25\%$\\
 \midrule
 CNN & \textbf{ 0.930} & 0.913 & $\blacktriangledown \textbf{2\%}$ & 0.663  & \textbf{0.564} & $\blacktriangledown \textbf{15\%}$ & 0.775 & \textbf{0.696} & $\blacktriangledown \textbf{11\%}$ & 0.888 & \textbf{0.846} & $\blacktriangledown \textbf{5\%}$\\
 
 HAN           &     0.871       & 0.912          & $\blacktriangle 5\%$          & \textbf{0.818}                    & 0.561  & $\blacktriangledown 31\%$ & \textbf{0.844} & 0.695 & $\blacktriangledown 18\%$ & \textbf{0.889} & 0.845 & $\blacktriangledown \textbf{5\%}$\\
\bottomrule
\end{tabular}
%
\vspace*{0.15cm}\caption*{\hspace*{-25pt}\begin{minipage}{355pt}
                Table 3: Temporal stability experiment. {\normalfont Results obtained by evaluating on the Wikipedia derived dataset from 2018 by either: models trained on Wikipedia data from 2018 or 2009. Trained on data from the same time frame, the neural models show a slight advantage over the lexical models. Most noticeable however is the drop in performance by the lexical models when trained on older data in terms of Recall and therefore also in terms of F1-score.}\end{minipage}}
\end{table}
\endgroup
  \end{minipage} \quad
  \begin{minipage}{.23\textwidth}
      \hspace*{-0.38cm}\vspace*{3.5cm}{\includegraphics[width=1.3\linewidth]{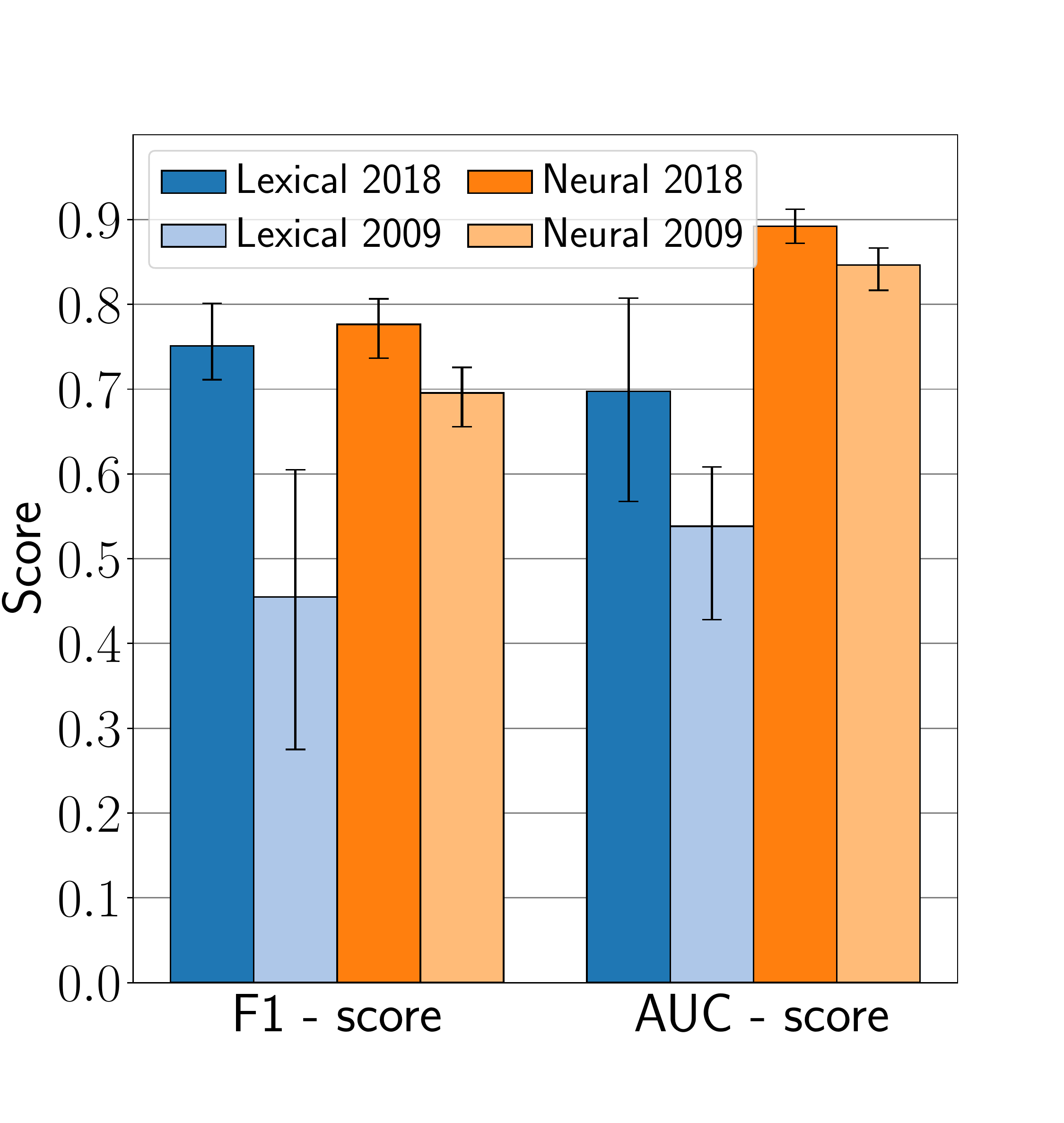}}
      \label{crosstime-fig}
      \captionsetup{width=1.1\linewidth}
      \vspace*{-4.5cm}\caption{Average F1 and AUC score of aggregated results for all lexical and neural models.}
  \end{minipage}
\end{figure*}

\section{Results}

\subsection{Comparison of results with previous work}
Table \ref{comparison} shows the relative performance of the neural models compared to previous controversy detection methods, evaluated on the Clueweb09 derived dataset of \cite{Dori13} and trained on the Wikipedia data from the same time frame. The TILE-Clique matching model outperforms all other models on Precision although this difference is not significant compared to the neural approaches. Similarly, the language model trained on the DBPedia dataset outperforms other models on Recall but shows no significant difference compared to the CNN model. Notably, the neural approaches show comparable results to the TILE-Clique model in terms of F1, demonstrating a balanced performance in terms of Precision and Recall. Furthermore, the CNN model shows a significant improvement compared to the other non neural baselines in terms of the AUC value (p < 0.05).

\begin{table}[H]
\centering
\setlength{\tabcolsep}{5.5pt} 
\caption{Comparison of results with previous work}
\label{comparison}
\begin{tabular}{lcccc}
\toprule
 Model                       &  Precision            & $\ $Recall$\ \ $           & $\ \ \ $F1$\ \ \ $                         & $\ \ $AUC$\ \ $ \\ \midrule
 Sentiment\_polyglot         & 0.448          & 0.392          & 0.418                    & 0.612 \\
 TfIdf-SVM                   & 0.581          & 0.208          & 0.306                    & 0.740 \\
 TILE\_Clique \cite{jang16}  & \textbf{0.710} & 0.720          & 0.714                    & 0.780 \\
 LM-DBPedia \cite{jang16b}   & 0.415          & \textbf{0.886} & 0.566                    & 0.730 \\
 LM-wiki                     & 0.359          & 0.808          & 0.497                    & 0.579 \\ \midrule
 CNN                         & 0.627          & 0.840          & \textbf{0.718}                    & \textbf{0.835} \\
 
 HAN                         & 0.632          & 0.745          & 0.684                    & 0.823 \\
\bottomrule
\end{tabular}
\end{table}

\setcounter{table}{3}

\subsection{Robustness of the model across time}

Controversy is expected to change over time. Some issues become controversial, others cease to be so. To investigate robustness of controversy detection models with respect to changes over time, we evaluate model performance in two variants: trained and tested on 2018, or trained on the 2009 Wikipedia data and tested on the 2018 Wikipedia data. Table 3 shows the results for each of the text-based detection models. 


\textbf{Within} year, the hierarchical attention model (HAN) outperforms all other models on Recall, F1 and AUC, losing Precision to the CNN and SVM models. However, our main interest is the robustness when a model is trained on a different year (2009) than the test set (2018). These \textbf{between} year experiments show a superior score for the HAN model compared to the non-neural models on Recall, and show significant improvements on F1 (p < 0.05) and AUC (p < 0.05), losing only to the SVM model on Precision (non significantly). In terms of robustness, we can also take the percentage change between the within year and between year experiment into account (were smaller absolute changes are preferable), shown by the delta values. With regard to temporal sensitivity, the CNN shows the least change across all four metrics. In Figure 1, we show the pooled results for the lexical and neural models to illustrate the overall increase in robustness by neural approaches. 

Interestingly, the SVM and HAN model show some unexpected improvement with regard to Precision when applied to unseen timeframes. For both models, this increase in Precision is offset by a greater loss in Recall, which seems to indicate both models `memorize` the controversial topics in a given timeframe instead of the controversial \textit{language}. Overall, the neural approaches seem to compare favorably in terms of cross-temporal stability.

\subsection{Robustness of the model across topics}

\begin{table}[H]
\centering
\setlength{\tabcolsep}{9pt} 
\caption{Cross-topic stability experiment. \textmd{Metrics are averaged across 10 leave-on-out topic folds.}}
\label{topic-table}
\begin{tabular}{lcccc}
\toprule
 Model                       &  Precision            & $\ $Recall$\ \ $           & $\ \ \ $F1$\ \ \ $                         & $\ \ $AUC$\ \ $ \\ 
 \midrule
 TfIdf-SVM  & 0.793          & 0.575 & 0.661                    & 0.829 \\
 LM  & 0.512          & \textbf{0.816} & 0.629                    & 0.633 \\
 \midrule
 CNN                         & \textbf{0.840}          & 0.569          & 0.670                    & \textbf{0.842} \\
 HAN                         & 0.799          & 0.716          & \textbf{0.753}                    & 0.840 \\
\bottomrule
\end{tabular}
\end{table}

To evaluate robustness towards unseen topics, 10-fold cross validation was used on the top ten largest topics present in the Wikipedia dataset in a leave-one-out fashion. The results are shown in table 4. In line with previous results, the language model scores best on Recall, beating all other models with a significant difference (p < 0.01). However in balancing Recall with Precision, the HAN model scores best, significantly outperforming both lexical models in F1 score (p < 0.05). Overall, when grouping together all neural and lexical results, the neural methods outperform the lexical models in Precision (p < 0.01), F1 (p < 0.05) and AUC (p < 0.01) with no significant difference found on the overall Recall scores. These results indicate that neural methods seem better able to generalize to unseen topics. 


\subsection{Robustness of the model across domains}
Most work on controversy has looked into using existing knowledge bases as a source of controversy information \cite{Dori13, jang16}. In this paper, we focus on text-based classification methods that do not aim to explicitly link general web pages to their knowledge-base counterparts. Therefore, we are interested in the ability of neural models to generalize beyond their training context. In addition to testing across time and topics, we also investigate robustness to changes in domain. By training only on Wikipedia data, and evaluating only on general web-pages, we look at the ability of the four methods to deal with out-of-domain documents. 

The hierarchical attention network shows significantly better results (p < 0.05) compared to all other models on F1. Both neural models also outperform both language models on AUC significantly (p < 0.05). Precision and Recall are more mixed, with the CNN and SVM outperforming the HAN on Precision and the language model -again- performing best in terms of Recall. Together, the neural methods seem to work best on three out of the four metrics.

\begin{table}[H]
\centering
\setlength{\tabcolsep}{9pt} 
\caption{Cross-domain stability experiment. \textmd{Metrics are based on models trained on Wikipedia data and tested on general web pages.}}
\label{domain-table}
\begin{tabular}{lcccc}
\toprule
 Model                       &  Precision            & $\ $Recall$\ \ $           & $\ \ \ $F1$\ \ \ $                         & $\ \ $AUC$\ \ $ \\
 \midrule
  TfIdf-SVM  & 0.718          & 0.361 & 0.480                    & 0.638 \\
 LM  & 0.392          & \textbf{0.826} & 0.532                    & 0.573 \\
 \midrule
 CNN                         & \textbf{0.743}          & 0.394          & 0.514                    & 0.755 \\
 HAN                         & 0.700          & 0.604          & \textbf{0.645}                    & \textbf{0.789} \\
\bottomrule
\end{tabular}
\end{table}

\subsection{Human agreement}

Lastly, we examine model performance with respect to human annotation using the human annotated dataset of \cite{Dori13}. We assume that models that perform similarly to human annotators are preferable. In Table \ref{human}, we present three Spearman correlation metrics to express model congruence with human annotations. Mean annotation expresses the correlation of model error rates with the controversy values attributed to a web page by human annotators, with positive values expressing greater error rates on controversial, and negative expressing higher error rates on non-controversial pages. Here, the HAN shows most unbiased (closest to zero) performance. 

Certainty is the distance of human annotations to the midpoint of the four-point controversy scale, i.e. a score between 0 and 2.5 that expresses how sure annotators are of document (non)controversy. Here, the HAN shows errors most strongly negatively correlated to the certainty of annotators. Finally, annotators disagree on the controversy of some documents, expressed as the standard deviation of their controversy annotations. Again, the HAN model seems preferable, as it's errors are most strongly correlated to annotator disagreement. Overall, the neural methods have less biased performance in relation to (non)controversial documents, correlate more strongly with the certainty of human annotators and are susceptible to errors in similar conditions as when annotators disagree. 

\begin{table}[H]
\setlength{\tabcolsep}{6pt} 
\caption{Spearman's correlations for estimated probability distance from true label. \textmd{Mean controversy: Average annotator score, certainty: distance from controversy annotation-scale midpoint, disagreement: standard deviation of annotations. Only pages with at least 3 annotations included to ensure sensible agreement metrics, N=128, bolded scores are preferable.}}
\begin{tabular}{lccc}
\toprule
\label{human}
Model &  mean annotation &  certainty &  disagreement  \\
\midrule
TfIdf-SVM    &            -0.540 &     -0.238 &           0.144 \\
LM-DBPedia     &             0.633 &     -0.172 &          -0.023 \\
\midrule
CNN    &             0.348 &     -0.314 &           0.138 \\
HAN    &             \textbf{0.277} &     \textbf{-0.390} &           \textbf{0.207} \\
\bottomrule
\end{tabular}
\end{table}

\section{Conclusion}

Controversy detection is a hard task, as it forms a latent concept sensitive to vocabulary gaps between topics and vocabulary shifts over time. We analysed the performance of language model, SVM, CNN and HAN models on different tasks. 

\textbf{First}, we have demonstrated that neural methods perform as state-of-the-art tools in controversy detection on the ClueWeb09 \cite{dori15} based testset, even beating matching models. \textbf{Second}, we investigated temporal stability, and demonstrated neural -and especially CNN- robustness in terms of Recall, F1 and AUC performance and stability with train and test sets that are 9 years apart. \textbf{Thirdly}, we show that CNN and HAN models outperform the SVM and LM baselines on Precision, F1 and AUC when tested on held-out-topics. \textbf{Fourthly}, we show that neural methods are better able to generalize from Wikipedia pages to unseen general web pages in terms of Precision, F1 and AUC. \textbf{Lastly}, neural methods seem better in line with human annotators with regard to certainty and disagreement.

\end{document}